# State Space Modeling of Inverter Based Microgrids Considering Distributed Secondary Voltage Control

Ali Dehghan Banadaki[1], *Student Member, IEEE*,
Farideh Doost Mohammadi[1], *Student Memebr, IEEE*, Ali Feliachi[1], *Senior Member, IEEE*
Lane Department of Computer Science and Electrical Engineering,
West Virginia University, Morgantown, WV 26505, USA

*Abstract*—Appropriate control of high penetration renewable energies in power systems requires a complete modeling of the system. In this paper, a comprehensive state space modeling of voltage source inverters, networks and loads are studied. We have modeled a secondary voltage control that utilizes the nominal set points of the output voltage as the input for the controller design. A distributed control algorithm only requiring a sparse communication between neighboring distributed generators (DGs) is used. The proposed secondary voltage model is applied on a microgrid with three DGs and two loads to verify the results.

## I. INTRODUCTION:

Microgrids (MGs) are a smaller scale of power systems that integrate high penetration of renewable energies and energy storage systems (ESSs) into the grid [1-6]. Power electronic converters are used as the interface between renewable energies such as solar power, wind power and fuel cells and the grid [7- 9]. Utilizing renewable energies in MGs as well as the flexibility of MGs in connecting and disconnecting to/from the grid is counted as two main advantages of MGs over the conventional power systems.

MGs can operate either in a grid connected mode or in a stand-alone mode [10, 11]. In the grid-connected mode, DGs are utilized to supply power for the MG while the voltage and the frequency of the system are controlled by the grid [12-14]. On the other hand, in a stand-alone mode, DGs control the voltage and the frequency of the system as well as meet the balance between generation and demand [15- 18].

A typical hierarchical control structure for controlling the MGs are divided into the three categories, namely primary, secondary and tertiary control levels [19, 20]. In a primary control level, droop controllers are utilizing a local negative feedback to control load disturbances in the system. The basic idea comes from synchronous generators that share any increase in the loads based on their ratings. The instantaneous increase in the power of demand will be compensated from the mechanical power of the rotor. The similar idea is used in the exciter control of synchronous machines (i.e. the voltage drops when the reactive power is increasing). These two characteristics are implemented in power electronic MGs via two droop equations [21]. However, primary control may not be able to bring back the voltage and the frequency of the system to its normal point hence secondary controller is applied to the system. The secondary controller can set the operating point based on the centralized controller. The global minimum can be found with the centralized method but disadvantages such as single point of failure and communication limitations limit its efficiency. Hence, distributed controller as an alternative method to the centralized controller is taking much more attraction [22, 23].

Previous dynamic analysis of standalone MGs has been carried out with some simplifying assumptions in the modeling of the system [24- 26]; therefore the analysis of the voltage control is not accurate on these models. In [27], a simple inductance L is modeled as the output filter while an RLC is needed to filter the high frequency modes. In [28], a single MG connected to an AC stiff bus is modeled; therefore the interaction among the inverters have not been considered. In [29], multiple ideal inverter based MGs has been modeled without modeling the fast response of the voltage control and the current control loops. In high switching frequency, dynamics of these two loops cannot be neglected. In [30], secondary voltage controller has been proposed on a MG with multiple DGs while the frequency effect on the inductance load and transmission lines is not considered. In [24], a comprehensive small signal model for inverter based MGs has been considered with considering the primary control loop only. In [31] secondary controller is proposed but only for the frequency control. In this paper, a secondary voltage controller has been proposed for a comprehensive small signal model of MGs (i.e. DGs, networks, and loads). In this method distributed voltage control is used that only needs limited information exchange among the immediate neighboring DGs [32, 33].

The rest of the paper is organized as follows: In section II, the state space model of a voltage control voltage source inverters (VCVSI) with considering the secondary voltage control is explained. Modeling of the network and load are also be given in this section. In section III, distributed voltage controller is explained. In section IV, the simulation for a MG with 3 DGs have been shown and lastly, the conclusion is given in section IV.

## II. STATE SPACE MODEL OF A VCVSIS:

The dynamical system of a VCVSI can be described with a nonlinear differential equation in (1):

$$\dot{x} = f(x(t), u(t), t) \qquad (1)$$

A linearized model of the form (1) around the nominal point can be found by considering the first order of Taylor Series expansion as in (2-3):

$$\Delta \dot{x}(t) = \dot{x}_0 + \Delta \dot{x}(t) \quad (2)$$

$$\dot{x}(t) = f(x_0(t), u_0(t), t) + A\Delta x(t) + B\Delta u(t) \quad (3)$$

where $A = \frac{\partial f}{\partial x}(x_0(t), u_0(t), t), B = \frac{\partial f}{\partial u}(x_0(t), u_0(t), t)$

Therefore by using equation (2, 3), state space model of a VCVSI can be found. VCVSIs are commonly used as an interface between renewable energies and the grid in the stand-alone mode. In this mode, the voltage and frequency of the system should be controlled via DGs of the system, hence VCVSIs are appropriate interfaces used in MGs. A VCVSI consists of a DC link, three leg inverters, an RLC filter and an RL coupling inductor. The controller part of the VCVSI consists of three parts including a power controller, a voltage controller and a current controller which have been shown in Fig. (1). A secondary voltage controller will be applied on the voltage set-point $v_n$ as it shown in Fig. (1). In the following part, each part of the controller is explained by following the method of ref. [24]:

### A. Power controller:

In a transmission system with a high transmission line ratio (X>>R), active power and reactive power are decoupled and can be controlled by the angle and magnitude of the bus voltage respectively. Therefore, the power flow can be managed via the voltage of each bus (i.e phase and magnitude). In conventional power systems, adjusting of active power is automatically done in the synchronous machines. Since the power from the mechanical part (i.e. rotor) will be transferred to the electrical part to compensate the demanding power; the frequency will be dropped based on the size of the rotor. Therefore the active power will be shared among the synchronous machines based on their ratings. The similar approach is implemented in the exciter of synchronous machines to control the reactive power of loads by changing the voltage magnitude. However, inverter based MGs cannot automatically do this since they do not have rotating parts (i.e. the inertia is null or too low). Power controller which is explained in the following has been used to implement these two characteristics in VCVSIs:

Each power controller has three blocks:

1) Power calculation block calculates the output power in the system via (4,5):

$$p = v_{od}i_{od} + v_{oq}i_{oq} \quad (4)$$

$$q = v_{od}i_{oq} - v_{oq}i_{od} \quad (5)$$

2) Low pass filter: due to the high switching impact of the inverter-based MGs on voltage and currents, a low pass filter is used in (6, 7) to filter the high frequency distortion of MGs:

$$P = (\omega_{co}/(s+\omega_{co}))p \quad (6)$$

$$Q = (\omega_{co}/(s+\omega_{co}))q \quad (7)$$

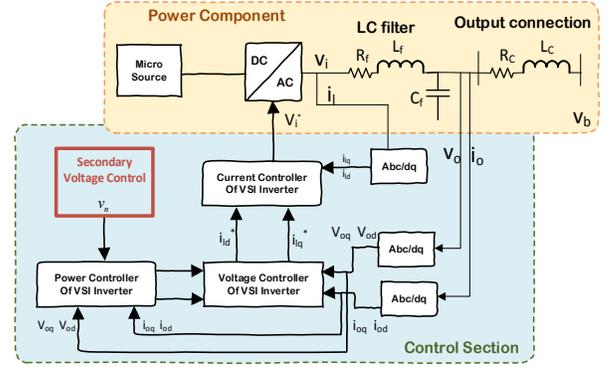

Figure 1: Block Diagram of VCVSI

3) Power Controller: Primary controller consists of two droop controllers that represent two negative output feedback controllers to share the active and reactive power based on the rating of the sources in MGs:

$$\omega = \omega_n - mP \quad (8)$$

$$v^*_{od} = V_n - nQ \quad (9)$$

$$v^*_{oq} = 0 \quad (10)$$

Where m and n are the droop gains that are calculated based on the rating of DGs as in (11, 12):

$$m = \frac{\Delta \omega_{max}}{P_{max}} \quad n = \frac{\Delta V_{od\,max}}{Q_{max}} \quad (11, 12)$$

In order to construct the complete model of a whole MG, everything should be transformed to a common reference frame. Here the angle of inverter 1 is assumed to be the reference, hence all other DGs' angles will be calculated from (13) and then can be used to transform the system to its common reference frame. It is obvious that this is similar to the transmission level where the angle of the slack bus is picked zero and all other angles will be found out based on the slack bus.

$$\delta = \int (\omega - \omega_{com}) \rightarrow \dot{\delta} = \omega - \omega_{com} \quad (13, 14)$$

By using Taylor Series, equations (6-10) can be written in a small signal form as in (15-16):

$$\begin{bmatrix} \Delta \dot{\delta} \\ \Delta \dot{P} \\ \Delta \dot{Q} \end{bmatrix} = A_p \begin{bmatrix} \Delta \delta \\ \Delta P \\ \Delta Q \end{bmatrix} + B_p \begin{bmatrix} \Delta i_{ldq} \\ \Delta v_{odq} \\ \Delta i_{odq} \end{bmatrix} \quad (15)$$

$$A_p = \begin{bmatrix} 0 & -m_p & 0 \\ 0 & -\omega_c & 0 \\ 0 & 0 & -\omega_c \end{bmatrix}, B_p = \begin{bmatrix} 0 & 0 & 0 & 0 & 0 & 0 \\ 0 & 0 & \omega_c I_{od} & \omega_c I_{oq} & \omega_c V_{od} & \omega_c V_{oq} \\ 0 & 0 & \omega_c I_{oq} & -\omega_c I_{od} & -\omega_c V_{oq} & \omega_c V_{od} \end{bmatrix}$$

$$\begin{bmatrix} \Delta \omega \\ \Delta v^*_{odq} \end{bmatrix} = \begin{bmatrix} C_{p\omega} \\ C_{pv} \end{bmatrix} \begin{bmatrix} \Delta \delta \\ \Delta P \\ \Delta Q \end{bmatrix} + \begin{bmatrix} 0 \\ B_{pvn} \end{bmatrix} [\Delta v_n] \quad (16)$$

$$C_{p\omega} = \begin{bmatrix} 0 & -m_p & 0 \end{bmatrix}, C_{pv} = \begin{bmatrix} 0 & 0 & -n_q \\ 0 & 0 & 0 \end{bmatrix}, B_{pvn} = \begin{bmatrix} 1 \\ 0 \end{bmatrix}$$

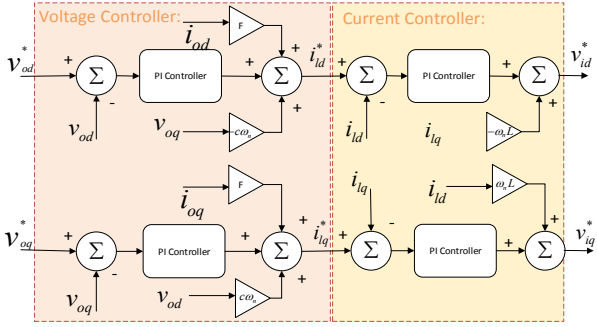

Figure 2. Voltage and Current Controller

B. *Voltage controller:* A voltage controller with standard PI controller considering the feedbacks and feedforwards is shown in Fig. (2) and formulated in (17-19).

$$\dot{\phi}_{dq} = v^*_{odq} - v_{odq} \tag{17}$$

$$i^*_{ld} = Fi_{od} - \omega_n C_f v_{oq} + K_{pv}(v^*_{od} - v_{od}) + K_{iv}\phi_d \tag{18}$$

$$i^*_{lq} = Fi_{oq} + \omega_n C_f v_{od} + K_{pv}(v^*_{oq} - v_{oq}) + K_{iv}\phi_q \tag{19}$$

Equation (17) can be rewritten by using equation (14) as in (20):

$$[\Delta\dot{\phi}_{dq}] = B_{V1}C_{pv}\begin{bmatrix}\Delta\delta\\ \Delta P\\ \Delta Q\end{bmatrix} + B_{V2}\begin{bmatrix}\Delta i_{ldq}\\ \Delta v_{odq}\\ \Delta i_{odq}\end{bmatrix} + B_{V1}B_{pvn}[\Delta v_n] \tag{20}$$

$$B_{V1} = \begin{bmatrix}1 & 0\\ 0 & 1\end{bmatrix}, B_{V2} = \begin{bmatrix}0 & 0 & -1 & 0 & 0 & 0\\ 0 & 0 & 0 & -1 & 0 & 0\end{bmatrix}$$

C. *Current Controller:* A Current controller which is shown in Fig. (2) can be used to find out the set points for the input voltage references:

$$\dot{\gamma}_{dq} = i^*_{ldq} - i_{ldq} \tag{21}$$

$$v^*_{id} = -\omega_n L_f i_{lq} + K_{pc}(i^*_{ld} - i_{ld}) + K_{ic}\gamma_d \tag{22}$$

$$v^*_{iq} = -\omega_n L_f i_{ld} + K_{pc}(i^*_{lq} - i_{lq}) + K_{ic}\gamma_q \tag{23}$$

Small signal model of equation (21) can be rewritten by using equations (9, 10, 18, 19) as in (24):

$$[\Delta\dot{\gamma}_{dq}] = [0][\Delta\gamma_{dq}] + B_{C1}D_{v1}C_{pv}\begin{bmatrix}\Delta\delta\\ \Delta P\\ \Delta Q\end{bmatrix} + (B_{C1}D_{v2} + B_{C2})\begin{bmatrix}\Delta i_{ldq}\\ \Delta v_{odq}\\ \Delta i_{odq}\end{bmatrix}$$

$$+ B_{C1}C_v[\Delta\phi_{dq}] + B_{C1}D_{v1}B_{pvn}[\Delta v_n] \tag{24}$$

D. *Output RLC filter and Coupling filter:* Nonlinear equations for the output filter regulating the high frequency distortions as well as the coupling filter are given in (25-30):

$$\frac{di_{ld}}{dt} = \frac{-r_f}{L_f}i_{ld} + \omega i_{lq} + \frac{1}{L_f}v_{id} - \frac{1}{L_f}v_{od} \tag{25}$$

$$\frac{di_{lq}}{dt} = \frac{-r_f}{L_f}i_{lq} - \omega i_{ld} + \frac{1}{L_f}v_{iq} - \frac{1}{L_f}v_{oq} \tag{26}$$

$$\frac{dv_{od}}{dt} = \omega v_{oq} + \frac{1}{C_f}i_{ld} - \frac{1}{C_f}i_{od} \tag{27}$$

$$\frac{dv_{oq}}{dt} = -\omega v_{od} + \frac{1}{C_f}i_{lq} - \frac{1}{C_f}i_{oq} \tag{28}$$

$$\frac{di_{od}}{dt} = \frac{-r_c}{L_c}i_{od} + \omega i_{oq} + \frac{1}{L_c}v_{od} - \frac{1}{L_c}v_{bd} \tag{29}$$

$$\frac{di_{oq}}{dt} = \frac{-r_c}{L_c}i_{oq} - \omega i_{od} + \frac{1}{L_c}v_{oq} - \frac{1}{L_c}v_{bq} \tag{30}$$

By using equation (18, 19) and equations (22, 23) with the assumption that $v_{idq} = v^*_{idq}$, state space modeling of this part can be derived as in (31):

$$\begin{bmatrix}\Delta\dot{i}_{ldq}\\ \Delta\dot{v}_{odq}\\ \Delta\dot{i}_{odq}\end{bmatrix} = (A_{LCL} + B_{LCL1}[D_{c1}D_{v2} + D_{c2}])\begin{bmatrix}\Delta i_{ldq}\\ \Delta v_{odq}\\ \Delta i_{odq}\end{bmatrix} + B_{LCL1}D_{c1}D_{v1}C_{pv}\begin{bmatrix}\Delta\delta\\ \Delta P\\ \Delta Q\end{bmatrix}$$

$$+ B_{LCL1}C_c[\Delta\gamma_{dq}] + B_{LCL1}D_{c1}C_v[\Delta\phi_{dq}] + B_{LCL2}[\Delta v_{bdq}] + B_{LCL3}[\Delta\omega]$$

$$+ B_{LCL1}D_{c1}D_{v1}B_{pvn}[\Delta v_n] \tag{31}$$

$$C_v = \begin{bmatrix}K_{iv} & 0\\ 0 & k_{iv}\end{bmatrix}, D_{v1} = \begin{bmatrix}K_{pv} & 0\\ 0 & K_{pv}\end{bmatrix}, D_{v2} = \begin{bmatrix}0 & 0 & -K_{pv} & -w_bC_f & F & 0\\ 0 & 0 & w_bC_f & -K_{pv} & 0 & F\end{bmatrix}$$

Equations of (15, 16, 20, 24, 31) can be used to find out the complete state space model for each inverter as in (32).

$$[\Delta\dot{x}_{invi}] = A_{INVi}[\Delta x_{INV}] + B_{INVi}[\Delta v_{bdq}] + B_{i\omega com}[\Delta\omega_{com}] + B_{vn}[\Delta v_n] \tag{32}$$

However, in order to combine the model of all of the inverters together, transformation equation (33) is needed to transform all of them to a common reference frame:

$$[f_{oDQ}] = [T_S][f_{odq}] \tag{33}$$

Where $T_S = \begin{bmatrix}\cos(\delta_i) & -\sin(\delta_i)\\ \sin(\delta_i) & \cos(\delta_i)\end{bmatrix}$. The small signal is given in (34):

$$[\Delta f_{oDQ}] = T_s(0)[\Delta f_{odq}] + T_C(0)[\Delta\delta] \tag{34}$$

Where $T_C = \begin{bmatrix}-F_{od}\sin(\delta_i) & -F_{oq}\cos(\delta_i)\\ F_{od}\cos(\delta_i) & -F_{oq}\sin(\delta_i)\end{bmatrix}$

By assuming $i_o$ (or $v_b$) as the output of the each inverter, then equation (34) can be used to transform the output of each inverter to the common reference frame:

$$[\Delta i_{oDQ}] = [T_S(0)][\Delta i_{odq}] + [T_C(0)][\Delta\delta] \tag{35}$$

On the other hand, $v_b$ (or $i_o$) is assumed to be the input signal to each inverter; therefore reverse transformation will be used as in (36):

$$[\Delta v_{bdq}] = [T_S(0)]^{-1}[\Delta v_{bDQ}] + [T_V(0)]^{-1}[\Delta\delta] \tag{36}$$

Where $T_V^{-1} = \begin{bmatrix}-V_{bD}\sin(\delta_i) + V_{bD}\cos(\delta_i)\\ -V_{bD}\cos(\delta_i) - V_{bD}\sin(\delta_i)\end{bmatrix}$

### E. Aggregated State Space Model of All Inverters:

Equations (35) and (36) can be used to find the state space model of each inverter in the common reference frame as in the equation (37):

$$\left[\Delta \dot{x}_{INVi}\right] = A_{INVi}\left[\Delta x_{INVi}\right] + B_{INVi}\left[\Delta v_{bDQi}\right] + B_{P\omega com}C_{INVcom}\left[\Delta x_{INVcom}\right]$$
$$+ B_{iVn}[\Delta v_n] \quad (37)$$

$$A_{INVi} = \begin{bmatrix} A_{Pi} & 0 & 0 & B_{Pi} \\ B_{V1i}C_{Pvi} & 0 & 0 & B_{V2i} \\ B_{C1i}D_{V1i}C_{Pvi} & B_{C1i}C_{Vi} & 0 & B_{C1i}D_{V2i}+B_{C2i} \\ B_{LCL1i}D_{C1i}D_{V1i}C_{Pvi}+B_{LCL3i}C_{Pvi} & B_{LCL1i}D_{C1i}C_{Vi} & B_{LCL1i}C_{Ci} & A_{LCLi}+B_{LCL1i}(D_{Ci}D_{V2i}+D_{C2i}) \\ +B_{LCL2i}[T_{Vi}^{-1} \; 0 \; 0] & & & \end{bmatrix}_{13*13}$$

$$B_{INVi} = \begin{bmatrix} 0 \\ 0 \\ 0 \\ B_{LCL2}T_S^{-1} \end{bmatrix}, B_{i\omega com} = \begin{bmatrix} B_{P\omega com} \\ 0 \\ 0 \\ 0 \end{bmatrix}, C_{INVcom} = \begin{bmatrix} C_{p\omega} \\ 0 \\ 0 \\ 0 \end{bmatrix}^T, B_{iVn} = \begin{bmatrix} 0 \\ B_{V1}B_{pvn} \\ B_{C1}D_{v1}B_{pvn} \\ B_{LCL1}D_{c1}D_{v1}B_{pvn} \end{bmatrix}$$

In a MG with several DGs, a combined inverter model that model the whole state space model is shown in (38):

$$\left[\Delta \dot{x}_{INV}\right] = A_{INV}\left[\Delta x_{INV}\right] + B_{INV}\left[\Delta v_{bDQ}\right] + B_{Vn}[\Delta v_n] \quad (38)$$

$$A_{INV} = \begin{bmatrix} A_{INVcom} + B_{1\omega com}C_{INVcom} & 0 & \dots & 0 \\ B_{2\omega com}C_{INVcom} & A_{INV2} & \dots & 0 \\ \dots & \dots & \dots & 0 \\ B_{s\omega com}C_{INVcom} & 0 & 0 & A_{INVs} \end{bmatrix}_{13s*13s}$$

$$B_{INV} = \begin{bmatrix} B_{INVcom} \\ B_{INV2} \\ \dots \\ B_{INVs} \end{bmatrix}, B_{Vn} = \begin{bmatrix} B_{iVN} \\ B_{iVN} \\ \dots \\ B_{iVN} \end{bmatrix}, C_{INVc} = \begin{bmatrix} C_{INVc1} & 0 & 0 & . \\ 0 & C_{INVc2} & 0 & . \\ . & . & . & . \\ . & . & . & C_{INVcs} \end{bmatrix}$$

### F. Network model:

A network consists of 's' inverters, 'n' lines and 'm' nodes. In a network with RL lines, the inductor current is the state variable and the equations for the whole network can be written as follows in (39):

$$\left[\Delta \dot{i}_{lineDQ}\right] = A_{NET}\left[\Delta i_{lineDQ}\right] + B_{1NET}\left[\Delta v_{bDQ}\right] + B_{2NET}\Delta \omega_{com} \quad (39)$$

$$A_{NET} = \begin{bmatrix} A_{NET1} & 0 & \dots & 0 \\ 0 & A_{NET2} & \dots & 0 \\ \dots & \dots & \dots & \dots \\ 0 & 0 & \dots & A_{NET1} \end{bmatrix}_{2n*2n}, B_{1NET} = \begin{bmatrix} B_{1NET1} \\ B_{1NET2} \\ \dots \\ B_{1NETn} \end{bmatrix}_{2n*1}, B_{2NET} = \begin{bmatrix} B_{2NET1} \\ B_{2NET2} \\ \dots \\ B_{2NETn} \end{bmatrix}_{2n*1}$$

$$B_{1NETi} = \begin{bmatrix} \dots & \frac{1}{L_i} & 0 & \dots & \frac{-1}{L_i} & 0 & \dots \\ \dots & 0 & \frac{1}{L_i} & \dots & 0 & \frac{-1}{L_i} & \dots \end{bmatrix}, A_{NETi} = \begin{bmatrix} \frac{-r_i}{L_i} & \omega_0 \\ \omega_0 & \frac{-r_i}{L_i} \end{bmatrix}, B_{2NETi} = \begin{bmatrix} I_{Qi} \\ -I_{Di} \end{bmatrix}$$

### G. Load Model:

RL load model that is considered in this paper is very similar to the Network model in equ. (39):

$$\left[\Delta \dot{i}_{loadQi}\right] = A_{load}\left[\Delta i_{loadDQ}\right] + B_{1LOAD}\left[\Delta v_{bDQ}\right] + B_{2LOAD}\Delta \omega_{com} \quad (40)$$

Where $A_{load}$, $B_{1LOAD}$ and $B_{2lOAD}$ are similar to the matrices $A_{NET}$, $B_{1NET}$, $B_{2NET}$ respectively as it is explained in [24].

### H. Complete Microgrid State Space Model:

State space modeling for inverters, lines, and loads have been defined in (38-40). In order to make sure that the node voltage is well-defined and numerical solution exists a virtual resistor $r_N$ is added between each node and the ground [24]. The value of $r_N$ is chosen to be high (1000 Ohms) so that it does not affect the results. The equations of each node can be written as in (41, 42):

$$v_{bDi} = r_N(i_{oDi} - i_{loadDi} + i_{lineDi,j}) \quad (41)$$
$$v_{bQi} = r_N(i_{oQi} - i_{loadQi} + i_{lineQi,j}) \quad (42)$$

Now for connecting the DGs to the Network and Loads, following equation can be used:

$$\left[\Delta v_{bDQ}\right] = R_N M_{INV}\left[\Delta i_{loadDQ}\right] + R_N M_{load}\left[\Delta i_{loadDQ}\right]$$
$$+ R_N M_{NET}\left[\Delta i_{lineDQ}\right] \quad (43)$$

Where matrix $M_{INV(2m*2s)}$ connects the inverters to the nodes. The entries of this matrix is one if DG-i is connected to the node j and zero otherwise. Matrix $M_{load(2m*2p)}$ connects the nodes to the loads with -1 as its entry and zero otherwise. Matrix $M_{NET(2m*2n)}$ maps the connecting lines onto the network nodes where it is 1 or -1 depending on entering the line current or leaving to the node. Finally $R_{N(2m*2m)}$ is a diagonal matrix which has $r_N$ as its diagonal entries.

Now the whole state space modeling can be written by using equations (38, 39, 40, 43) as in (44):

$$\begin{bmatrix} \Delta \dot{x}_{INV} \\ \Delta_{ilineDQ} \\ \Delta_{iloadDQ} \end{bmatrix} = A_{mg} \begin{bmatrix} \Delta x_{INV} \\ \Delta_{ilineDQ} \\ \Delta_{iloadDQ} \end{bmatrix} + B_{vn}\Delta V_n \quad (44)^1$$

## III. DISTRIBUTED SECONDARY VOLTAGE CONTROL:

An autonomous controller in primary level, stabilize the system and share the load based on the ratings of each inverter. The primary control level is automatic and it does not rely on communication. However, it might not be able to

---

$$A_{mg} = \begin{bmatrix} A_{INV} + B_{INV}R_N M_{INV}C_{INVc} & B_{INV}R_N M_{NET} & B_{INV}R_N M_{load} \\ B_{1NT}R_N M_{INV}C_{INVc} + B_{2NET}C_{INV\omega} & A_{NET} + B_{1NET}R_N M_{NET} & B_{1NET}R_N M_{load} \\ B_{1load}R_N M_{INV}C_{INVc} + B_{2LOAD}C_{INV\omega} & B_{1LOAD}R_N M_{NET} & A_{load} + B_{1LOAD}R_N M_{LOAD} \end{bmatrix}, B_{vn} = \begin{bmatrix} B_{Vn} \\ 0 \\ 0 \end{bmatrix}$$

bring back the voltages to their nominal values. Hence the secondary control level will be applied to the system.

In the secondary control level, the voltage set point (i.e. $v_n$ in equation (2)) will be adjusted to regulate the output voltage. A distributed cooperative tracker algorithm is proposed in [32]. In this algorithm, only the immediate neighbors will share the voltage output among each other. The reference voltage is just needed to be known by one of the DGs that is connected to any other DGs in the system directly or indirectly (i.e. communication system has a spanning tree). The differences between the neighboring voltages will be calculated as it is shown in equation (45) and Fig. (3) to change the $v_n$. In addition, only the root node find out the difference from the $v_{ref}$ (i.e. $g_i$=1 for the root node and zero for the others)

$$\Delta v_{ni} = c_{vi} \int (\sum_j a_{ij}(v_{odi} - v_{odj}) + g_i(v_{odi} - v_{ref}))dt \quad (45)$$

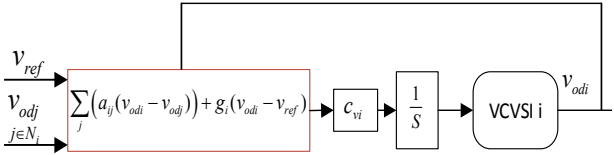

Figure 3: Distributed secondary voltage control

## IV. SIMULATION

A system with 3 DGs and 2 loads is studied here. The state space modeling of the system have been calculated based on the equation (53) and is implemented in Matlab Simulink. System parameters including controller parameters are shown in table I. System's initial condition is shown in table II and can be found from the load flow solution. However, the nonlinear model of the system can also be used to find the initial conditions [24]. At the time equal to 0.2 seconds, a load disturbance is added to the system. Then the secondary controller is activated at t=1 second to show how it improves the voltage deviation after the disturbance.

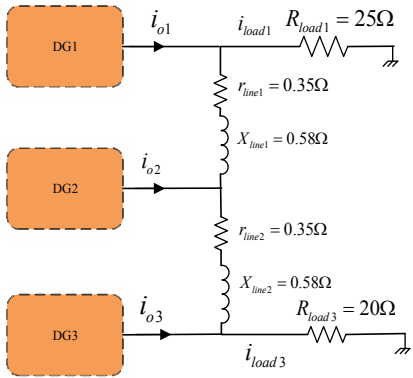

Figure 4: Test System

TABLE I: UNDERSTUDIED SYSTEM PARAMETERS

| Inverters Parameters | | | | | | | |
|---|---|---|---|---|---|---|---|
| $r_f$ | 0.1 Ω | $L_c$ | 0.35 mH | $K_{pc}$ | 10.5 | $m_p$ | 9.4e-5 |
| $L_f$ | 1.35 mH | $\omega_c$ | 31.41 | $K_{ic}$ | 16e3 | $n_q$ | 1.3e-3 |
| $C_f$ | 50 μF | $r_c$ | 0.03 Ω | $K_{pv}$ | 0.05 | $K_{iv}$ | 390 |
| $c_v$ | [-5 10 -10] | $g_1$ | 1 | $a_{12}$ | 1 | $a_{13}$ | 1 |

TABLE II. Initial Conditions [24]

| Par. | Value | Par. | Value |
|---|---|---|---|
| $V_{odq}$ | $\begin{bmatrix}380.8 & 381.8 & 380.4\\ 0 & 0 & 0\end{bmatrix}$ | $V_{bd}$ | $\begin{bmatrix}379.5 & 380.5 & 379\\ -6 & -6 & -5\end{bmatrix}$ |
| $I_{odq}$ | $\begin{bmatrix}11.4 & 11.4 & 11.4\\ 0.4 & -7.3 & -4.3\end{bmatrix}$ | $I_{ldq}$ | $\begin{bmatrix}11.4 & 11.4 & 11.4\\ -5.5 & -7.3 & -4.6\end{bmatrix}$ |
| $I_{line1dq}$ | $\begin{bmatrix}-3.8\\ 0.4\end{bmatrix}$ | $I_{line2dq}$ | $\begin{bmatrix}7.6\\ -1.3\end{bmatrix}$ |
| $\omega_0$ | 314 | $\delta_0$ | $[0 \quad 1.9e-3 \quad -0.0113]$ |

*1)* Scenario one: in the first scenario the voltage profile is regulated via an ideal distributed algorithm which has been explained in the previous section. It can be seen from Fig. 5 that the voltages are going back to their normal operating points by changing the voltage set points of each inverter.

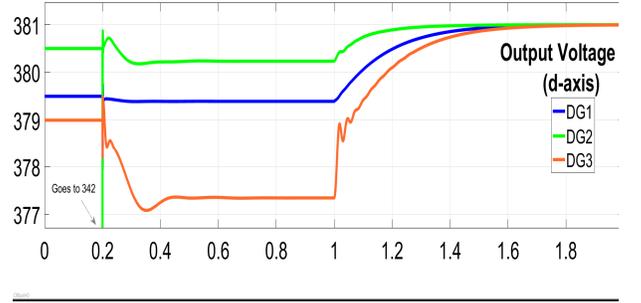

Figure 5: Output voltage profile

*2)* Scenario two: in the second scenario, a time delay of 0.1 seconds is added to the system. Although the system has more oscillations and needs more time to reach its steady state but it will be stable eventually. The result is shown is a Fig. 6.

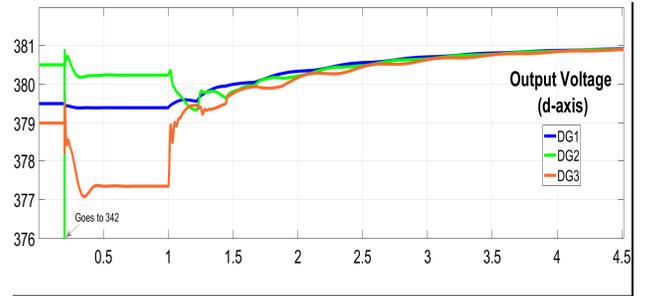

Figure 6: Output voltage profile with considering delay

## V. Conclusions

In this paper, a secondary voltage control has been proposed for a comprehensive state space modeling of an inverter-based MG which was introduced in [24]. A distributed control algorithm has been applied on the proposed state space model to verify the model. The results show that the system achieves secondary voltage control via the proposed state space model.